\documentclass[]{iopart}
\usepackage{graphics}
\usepackage{cite}
\newcommand{\ud}{\mathrm{d}}
\begin{document}
\title{Two-gap superconductivity: interband interaction in the role of external field}
\author{Artjom Vargunin, K\"ullike R\"ago and Teet \"Ord}
\ead{teet.ord@ut.ee}
\address{Institute of Physics, University of Tartu, 4 T\"ahe Street,
51010 Tartu, Estonia}

\begin{abstract}We show that weak interband pair-transfer interaction affects weaker-superconductivity component of a two-gap system as an external field associated with the order parameter, similarly to an external magnetic field for ferromagnetics or an external electric field for ferroelectrics.
\end{abstract}


\section{Introduction} Although low-temperature properties of some metals  \cite{metals0a,metals0b,metals0c,metals0d} and doped SrTiO$_3$ \cite{SrTiO0} were found to follow the Bardeen-Cooper-Schrieffer theory with overlapping energy bands \cite{Suhl,Moskalenko}, multi-gap superconductivity was treated rather exotic in the last century. Nowadays, the physics of cuprates \cite{cuprates0a,cuprates0b,cuprates0c}, MgB$_2$ \cite{MgB0a,MgB0b,MgB0c,MgB0d} and pnictides
\cite{pnictides0a,pnictides0b,pnictides0c,pnictides0d,pnictides0e},
among others, draws enormous attention to that scenario.

An interaction between superconductivity components plays the crucial role for two-band systems. Interband coupling raises critical temperature of the condensate. That pairing causes the penetration of the Cooper pairs from the stronger band (i.e. band with stronger superconductivity) into weaker one inducing there additional contribution to superconductivity. This peculiarity becomes important in the temperature region where stronger band is intrinsically superconducting (active), but weaker one is not yet, and results in the simultaneous vanishing of the band gaps in spite of distinct intraband critical points.

An effect of interband pairing is most intriguing in the situation when joint superconducting state is close to the splitting into two independent condensates. In this case the memory about intrinsic criticality of the weaker band becomes very pronounced and visible in the temperature scale as an additional maximum or a kink in the heat capacity curve \cite{capacity1,capacity2,capacity3,capacity4,capacity5,me2}, coherence lengths \cite{babaev1,litak2012,ord1}, relaxation times \cite{ord2}, conductivity \cite{VSi1} and superfluid density data \cite{Vsi2,FeSe1}. At that, there appears an inflection point in the temperature behaviour of the weaker-band gap which was observed in a number of two-gap materials \cite{SrTiO0,SrTiO1,NbSe,RNiBC1,RNiBC2,RNiBC3,LaC,MgAlB1}. Here we mention heavy fermion \cite{UNiAl} and multilayered  \cite{Cu1234a,Cu1234b} systems which also demonstrate distinct superconductivity gaps opening fully at different temperatures. 

The evolution of weaker-band superconducting instability with increase of interband coupling indicates close analogy with the modifications of the phase transition anomaly under applied external field \cite{landau}. Generally, that field rounds the singularities of various characteristics appearing at the phase transition point and locates them into somewhat shifted positions. Moreover, these tendencies are common within certain universality class with scaling behaviour in the known form. The analogy between interband pairing and external field can essentially enrich an understanding of two-gap physics. In particular, it may be of importance for type-1.5 superconductivity \cite{babaev2} for which the proximity to intrinsic critical point of the weaker band plays in favour.

The behaviour of the gap healing length, analysed recently for two-band scenario, reveals a peak close to the superconducting phase transition point of the weaker band taken as independent system. The magnitude of that peak scales with interband interaction constant with an exponent $-\frac{1}{3}$ \cite{shanenko}. Since the same exponent appears in the Landau mean-field theory for correlation length, it may signal the correctness of the analogy between interband coupling and external field. However, justification of this similarity calls for the analysis of other characteristics of a system. Moreover, one should be cautious due to discrepancy between spatial scales of coherency and recovery of the gap functions in a two-band superconductor \cite{babaev3}.

In the present contribution we perform a detailed examination of the
smearing of intraband superconducting instability by interband pairing
seen in the temperature behaviour of thermodynamic functions. From the
analysis carried out it follows that weak interband interaction
affects the various quantities related to the weaker band as an external
field associated with the order parameter.

\section{Thermodynamics of a two-gap system} We start with the Hamiltonian of a homogeneous two-band superconductor with intra- and interband pair-transfer interactions,
\begin{eqnarray}\label{e1}
&&H=\sum_{\alpha\mathbf{k}s}\tilde{\epsilon}_\alpha(\mathbf{k})a^+_{\alpha\mathbf{k}s}a_{\alpha\mathbf{k}s}\nonumber\\
&&-\frac{1}{V}\sum_{\alpha\alpha^\prime}\sum_{\mathbf{k}\mathbf{k^\prime}}W_{\alpha\alpha^{\prime}}a^+_{\alpha\mathbf{k}\uparrow}a^+_{\alpha-\mathbf{k}\downarrow}a_{\alpha^\prime-\mathbf{k^\prime}\downarrow}a_{\alpha^\prime\mathbf{k^\prime}\uparrow},\qquad
\end{eqnarray}
where $\tilde{\epsilon}_\alpha=\epsilon_\alpha-\mu$ is the electron
energy in the band $\alpha=1,2$ relative to the chemical potential $\mu$; $V$ is the volume of superconductor and $W_{\alpha\alpha^{\prime}}$ are the
matrix elements of intraband attraction ($W_{\alpha\alpha}>0$) or interband
($\alpha\neq\alpha^\prime$) interaction. It is supposed that the
chemical potential is located in the region of the bands
overlapping. We assume that (effective) electron-electron
interactions are nonzero only in the layer $\mu\pm\hbar\omega_\mathrm{D}$ and that the interaction constants are independent on electron wave
vector in this layer. We also take $W_{12}=W_{21}$.

To calculate the partition function
$Z=\mathrm{Sp}\exp\left(\frac{-H}{k_\mathrm{B}T}\right)$ we use the Hubbard-Stratonovich transformation \cite{Hubbard,Stratonovich} by means of which one linearizes and diagonalizes an exponent in $Z$ by
introducing complex integration variables. For sufficiently small
interband interaction $W^2=W_{11}W_{22}-W_{12}^2>0$ the static path
approximation reads as
\begin{eqnarray}\label{e2}
&&Z=\left(\frac{V}{\pi
k_BTW}\right)^2\int\limits_{-\infty}^{\infty}\exp\left(-\frac{V\tilde{f}}{k_\mathrm{B}T}\right)\ud\delta^\prime_{1}\ud\delta^{\prime\prime}_{1}\ud\delta^\prime_{2}\ud\delta^{\prime\prime}_{2}
,\qquad\\
&&\tilde{f}=\sum_{\alpha}\Bigg[\frac{1}{V}\sum_\mathbf{k}\left
(\tilde{\epsilon}_\alpha(\mathbf{k})-
2k_{B}T\ln\bigg(2\mathrm{ch}\frac{\tilde{E}_{\alpha}(\mathbf{k})}{2k_\mathrm{B}T}\bigg)\right)\nonumber\\
&&+\frac{W_{3-\alpha,3-\alpha}}{W^2}|\delta_\alpha|^2-\frac{W_{\alpha,3-\alpha}}{W^2}(\delta^\prime_\alpha\delta^\prime_{3-\alpha}+\delta^{\prime\prime}_\alpha\delta^{\prime\prime}_{3-\alpha})\Bigg].\label{e3}
\end{eqnarray}
Here the integration variables $\delta^\prime_\alpha$ and
$\delta^{\prime\prime}_{\alpha}$ are treated as real and
imaginary parts of the non-equilibrium complex order parameters
$\delta_\alpha$, and
\begin{equation}\label{e4}
\tilde{E}_{\alpha}(\mathbf{k})=\sqrt{\Big(\tilde{\epsilon}_\alpha(\mathbf{k})-\frac{W_{\alpha\alpha}}{2V}\Big)^2+|\delta_\alpha|^2}.
\end{equation}

Next we find the equilibrium free energy density $f=-k_\mathrm{B}T\frac{\ln Z}{V}$ for a macroscopic superconductor ($V\to\infty$). First, we perform an integration in $Z$ over the phases of non-equilibrium order parameters  and then go to infinite volume. In this process one obtains the integration over wave vector $\mathbf{k}$ instead of summation in $\tilde{f}$, which we replace with the integration over energy. Free energy density $f$ becomes
\begin{eqnarray}\label{e5}
&&f=f_\mathrm{n}+\sum_\alpha\Bigg[-4k_\mathrm{B}T\rho_\alpha\int\limits_{0}^{\hbar\omega_\mathrm{D}}
\ln\frac{\mathrm{ch}\frac{E_\alpha(\tilde\epsilon_\alpha)}{2k_\mathrm{B}T}}
{\mathrm{ch}\frac{\tilde{\epsilon}_\alpha}{2k_\mathrm{B}T}}\ud\tilde{\epsilon}_\alpha\nonumber\\
&&+\frac{W_{3-\alpha,3-\alpha}}{W^2}\Delta_\alpha^2-\frac{|W_{\alpha,3-\alpha}|}{W^2}\Delta_\alpha\Delta_{3-\alpha}\Bigg].
\end{eqnarray}
Here $E_\alpha(x)=\sqrt{x^2+\Delta_\alpha^2}$, where $\Delta_\alpha$ is the modulus of equilibrium order parameter, i.e. the value of $|\delta_\alpha|$ which minimizes non-equilibrium free energy density $\tilde{f}$. The quantity $f_\mathrm{n}$ corresponds to the free energy density in
the absence of superconductivity, and $\rho_\alpha$ is the density of electron states taken to be constant in the narrow integration layer around the Fermi level.

The minimization of non-equilibrium free energy leads to the equations for equilibrium order parameters
\begin{equation}\label{e6}
\frac{W_{3-\alpha,3-\alpha}}{W^2}\Delta_\alpha-\rho_\alpha\Delta_\alpha\int\limits_{0}^{\hbar\omega_\mathrm{D}}
\mathrm{th}\frac{E_\alpha(\tilde\epsilon)}{2k_\mathrm{B}T}\frac{\ud\tilde\epsilon}{E_\alpha(\tilde\epsilon)}=\frac{|W_{\alpha,3-\alpha}|}{W^2}\Delta_{3-\alpha}.
\end{equation}
If interband interaction is absent, the latter system
splits into two independent equations which describe intrinsic
superconductivity in the bands. The corresponding critical
temperatures equal
$T_{\mathrm{c}\alpha}=1.13\frac{\hbar\omega_\mathrm{D}}{k_\mathrm{B}}e^{-\frac{1}{\rho_\alpha W_{\alpha\alpha}}}$. Below we assume that $\alpha=1$ corresponds to the stronger band, i.e. $T_{\mathrm{c}1}>T_{\mathrm{c}2}$. If interband coupling is present the critical temperatures $T_{\mathrm{c}\alpha}$ transform into
\begin{equation}\label{e7}
T_{\mathrm{c}\mp}=1.13\frac{\hbar\omega_\mathrm{D}}{k_\mathrm{B}}e^{-\frac{2}{\rho_1W_{11}
+\rho_2W_{22}\pm\sqrt{(\rho_1W_{11}-\rho_2W_{22})^2+4\rho_1\rho_2W_{12}^2}}},
\end{equation}
where $T_{\mathrm{c}-}>T_{\mathrm{c}+}$. With $|W_{12}|$ increase the temperature $T_{\mathrm{c}-}$ increases and $T_{\mathrm{c}+}$
decreases approaching zero as $W\to0$. For $W_{12}\to0$ one has
$T_{\mathrm{c}-}\to T_{\mathrm{c}1}$ and $T_{\mathrm{c}+}\to
T_{\mathrm{c}2}$. Note that for a superconductor with interacting bands there is only one phase transition temperature $T_\mathrm{c}=T_{\mathrm{c}-}$. However, the point $T_{\mathrm{c}+}$ is also meaningful, because below $T_{\mathrm{c}+}$ the metastable superconducting states or saddle-points of non-equilibrium free energy appear \cite{Soda,me1}. These peculiarities can be reflected in the behaviour of superconducting fluctuations.

Thermodynamics of the system related to superconductivity is entirely described by excess free energy density $\Delta f=f-f_\mathrm{n}$ and its temperature derivatives, e.g. excess specific entropy $\Delta s=-\Delta f^\prime$ and specific heat capacity $\Delta c=-T\Delta f^{\prime\prime}$. For the degenerate electron gas one has $s_\mathrm{n}=c_\mathrm{n}=\gamma_\mathrm{S}T$, where $\gamma_\mathrm{S}=\sum_\alpha\gamma_{\mathrm{S}\alpha}$
is the Sommerfeld constant and $\gamma_{\mathrm{S}\alpha}=\frac{2}{3}\pi^2k_\mathrm{B}^2\rho_\alpha$. Therefore, the extrema and inflection points of $\Delta s$ coincide in the temperature scale with zeros and extrema of $\frac{\Delta c}{c_\mathrm{n}}$, correspondingly. 

In a two-band superconductor thermodynamic functions may be represented by means of additive contributions from the relevant bands, i.e.  $\Delta f=\sum_\alpha\Delta f_\alpha$, $\Delta s=\sum_\alpha\Delta s_\alpha$ and $\Delta c=\sum_\alpha\Delta c_\alpha$. However, due to interband pairing, the relations between band contributions to the thermodynamic functions may be dissimilar to those known for non-interacting subsystems.

\section{Critical exponents}
Since thermodynamics driven by free energy (\ref{e5}) can be analysed only numerically, one should use certain restrictions to get analytic results. By starting with non-equilibrium free energy $\tilde{f}$ expanded in powers of $\delta_\alpha$, we obtain for a macroscopic superconductor
\begin{equation}\label{e8}
\Delta
f=\sum_\alpha\left( a_\alpha\Delta_\alpha^2+\frac{b_\alpha}{2}\Delta_\alpha^4-\gamma\Delta_\alpha\Delta_{3-\alpha}\right),
\end{equation}
where
\begin{equation}\label{e9}
a_\alpha=\frac{W_{3-\alpha,3-\alpha}}{W^2}-\rho_\alpha\ln\frac{1.13\hbar\omega_\mathrm{D}}{k_\mathrm{B}T},
\end{equation}
and $b_\alpha=\frac{0.11\rho_\alpha}{(k_\mathrm{B}T)^2}$, $\gamma=\frac{|W_{12}|}{W^2}$. This approximation works perfectly in the vicinity of critical point $T_\mathrm{c}=T_{\mathrm{c}-}$. However, as we do not expand the parameters $a_\alpha$ and $b_\alpha$ in powers of $T_\mathrm{c}-T$, we have at least qualitatively correct picture also  further off the critical temperature \cite{GL}. In particular, one ascertains numerically that corresponding gap equations
\begin{equation}\label{e10}
a_\alpha\Delta_\alpha+b_\alpha\Delta_\alpha^3=\gamma\Delta_{3-\alpha}
\end{equation}
describe a kink in the behaviour of smaller gap $\Delta_2(T)$ near $T_{\mathrm{c}+}$. Note that alternative approaches \cite{shanenko2} based on the expansion in powers of $T_\mathrm{c}-T$ cannot reproduce that kink  for tiny interband coupling.

Next we solve analytically the system (\ref{e10}) in the vicinity of $T_{\mathrm{c}+}$ for extremely small interband interactions. Corresponding equation for $\Delta_2(T)$ reads as 
\begin{equation}\label{e11}
\Delta_2^2\left(\left(\Delta_2^2-x\right)^3-y^3\right)=\frac{\gamma^2(\gamma^2-a_1a_2)}{b_1b_2^3},
\end{equation}
where $x=-\frac{a_2}{b_2}$ and $y^3=-\frac{a_1\gamma^2}{b_1b_2^2}$. As the condition $\gamma^2-a_1a_2=0$ determines the points $T_{\mathrm{c}\pm}$, we obtain $\Delta^2_{2+}=x_++y_+$. Here and elsewhere index $+$ stands for $T=T_{\mathrm{c}+}$. Next we expand these values in powers of normalized interband coupling $w=\left(\frac{W_{12}}{\rho_2W_{11}W_{22}}\right)^2$ by using the expansion forms $x_+=\sum_{n=0}^\infty x_+^{\{1+n\}}$ and $y_+=\sum_{n=0}^\infty y_+^{\{\frac{1}{3}+n\}}$ with summation over the contributions of different powers of $w$ denoted by upper indexes in curly brackets, for instance, $x_+^{\{n\}}\sim w^n$. We also introduce the quantities $A=\frac{\rho_2}{b_2}$ and $B=\frac{\gamma^2}{b_2^2}$ for which $A_+=\sum_{n=0}^\infty A_+^{\{n\}}$ and $B_+=\sum_{n=0}^\infty B_+^{\{1+n\}}$. The first terms of these expansions are given by
\begin{eqnarray}
&&x_+^{\{1\}}=A_+^{\{0\}}\frac{\rho_2}{\rho_1}t^{-1}w,\qquad y_+^{\{\frac{1}{3}\}}=A_+^{\{0\}}t^\frac{1}{3}w^\frac{1}{3},\nonumber\\
&&A_+^{\{0\}}=9.4\big(k_\mathrm{B}T_{\mathrm{c}2}\big)^2,\qquad B_+^{\{1\}}=A_+^{\{0\}2}w,\label{e12}
\end{eqnarray}
and $t=\ln\frac{T_{\mathrm{c}1}}{T_{\mathrm{c}2}}$. Note also that $\frac{T_{\mathrm{c}+}}{T_{\mathrm{c}2}}\approx1+o(w)$.

By differentiating Eq. (\ref{e11}) at $T_{\mathrm{c}+}$, we find the consecutive temperature derivatives of $\Delta_2$. For instance, in the lowest order in $w$ we have 
\begin{equation}\label{e13}
\Delta_{2+}=\sqrt{y_+^{\{\frac{1}{3}\}}},\qquad\frac{\Delta_{2+}^\prime}{\Delta_{2+}}=-\frac{A_+^{\{0\}}}{3T_{\mathrm{c}+}y_+^{\{\frac{1}{3}\}}},\label{ea4}
\end{equation}
etc. Usually $w$ power of the lowest order contributions to the consecutive derivatives at $T_{\mathrm{c}+}$ decreases by $\frac{1}{3}$ as order of the derivative increases, for instance, $\frac{\Delta_{2+}}{\Delta_{2+}^{\prime}}\sim w^\frac{1}{3}$. Sometimes, however, corresponding $w$ power decreases by $\frac{2}{3}$ due to extra vanishing of the terms. The latter takes place e.g. for $\Delta_{2+}^{\prime\prime}$ which results in $\frac{\Delta_{2+}^{\prime}}{\Delta_{2+}^{\prime\prime}}\sim w^0$ and $\frac{\Delta_{2+}^{\prime\prime}}{\Delta_{2+}^\mathrm{I\!I\!I}}\sim w^\frac{2}{3}$.

The temperature dependence of the smaller gap is given by the Taylor series
$\Delta_2(T)=\sum_{n=0}^\infty\Delta_{2+}^{(n)}\frac{(T-T_{\mathrm{c}+})^n}{n!}$, where index in parentheses denotes the order of the temperature derivative. The value $\Delta_2(T_{\mathrm{c}2})$ contains the sum of $w$ dependent contributions which become smaller and smaller as $n$ increases. For tiny $|W_{12}|$ this value is defined predominantly by $\Delta_{2+}$, or
\begin{equation}\label{e14}
\Delta_2(T_{\mathrm{c}2})\sim |W_{12}|^\frac{1}{3}.
\end{equation}
The "susceptibility" related to the "field" $|W_{12}|$ reads as
\begin{equation}\label{e15}
\frac{\partial\Delta_2(T_{\mathrm{c}2})}{\partial |W_{12}|}\sim |W_{12}|^{-\frac{2}{3}}.
\end{equation}

The weaker-band contribution to excess specific entropy is
\begin{eqnarray}\label{e16}
\Delta
s_2=-a_2^\prime\Delta_\alpha^2-\frac{b_2^\prime}{2}\Delta_2^4,
\end{eqnarray}
and $\Delta c_2=T\Delta s_2^\prime$. Similarly to the situation with the smaller gap, the values $\Delta s_2(T_{\mathrm{c}2})$ and $\Delta c_2(T_{\mathrm{c}2})$ are defined predominantly by $\Delta s_{2+}$ and $\Delta c_{2+}$, correspondingly, for tiny $|W_{12}|$. As a result,
\begin{eqnarray}\label{e17}
&&\Delta s_2(T_{\mathrm{c}2})\sim |W_{12}|^\frac{2}{3},\\
&&\Delta c_2(T_{\mathrm{c}2})\sim |W_{12}|^0.\label{e18}
\end{eqnarray}

\begin{figure*}
\resizebox{1.1\columnwidth}{!}{
\includegraphics{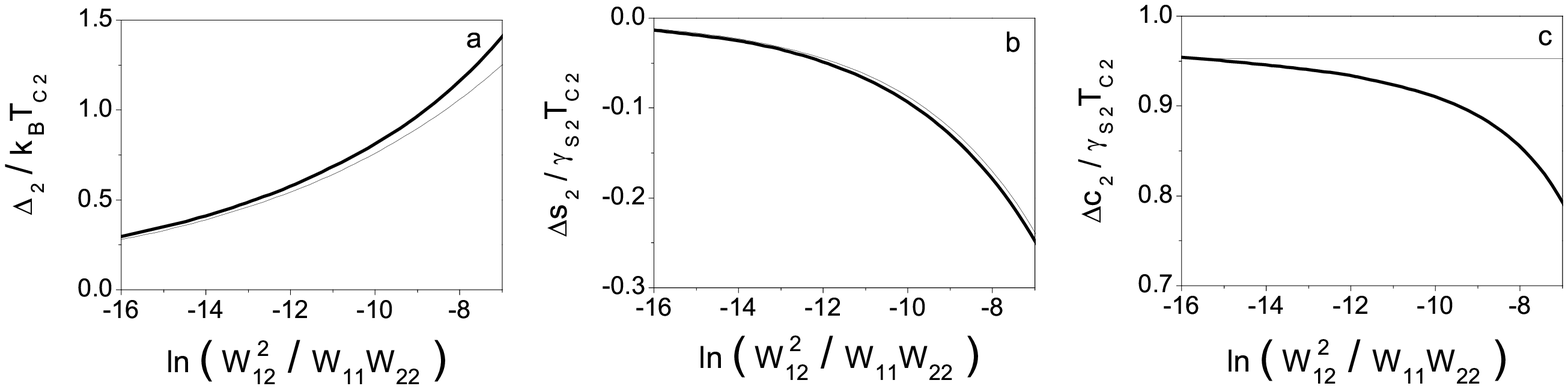}}\vspace{-6cm}
\caption{The plots of the weaker-band gap (a), corresponding excess specific entropy (b) and normalized excess specific heat capacity (c) vs interband interaction constant at $T_{\mathrm{c2}}$. The thick curves represent thermodynamics based on Eqs. (\ref{e5}) and (\ref{e6}), while thin ones correspond to the dependencies (\ref{e14}), (\ref{e17}) and (\ref{e18}) found for tiny $|W_{12}|$.}\label{f1}
\end{figure*}

Next we compare these dependencies with thermodynamics based on free energy (\ref{e5}) and gap equations (\ref{e6}). For illustration we consider the following set of intraband parameters: $\rho_{1,2}=(1,0.9)
(\mathrm{eV}\cdot\mathrm{cell})^{-1}$, $W_{11,22}=0.3\mathrm{\
eV}\cdot\mathrm{cell}$ and $\mathrm{cell}=0.1\mathrm{\ nm}^3$. In
this case $T_{\mathrm{c2}}=0.69T_{\mathrm{c1}}$. Fig. \ref{f1} shows that restrictions made do no harm the description of the weaker-superconductivity component.

In the Landau mean-field theory of criticality an applied external field $h$ introduces additional contribution $h\Delta_\mathrm{L}$ to the free energy expansion and forces the order parameter $\Delta_\mathrm{L}$ to change as $h^\frac{1}{\delta}$ ($\delta=3$), susceptibility as $h^{\frac{1}{\delta}-1}$, entropy as $h^\epsilon$ ($\epsilon=\frac{2}{3}$) and heat capacity as $h^{-\alpha_\mathrm{c}}$ ($\alpha_\mathrm{c}=0$). The weaker-band superconducting instability modifies with interband interaction constant in the same way, see Eqs. (\ref{e14}), (\ref{e15}), (\ref{e17}), and (\ref{e18}). Thus, interband coupling acts as external field governing the intrinsic criticality of weaker band. At that, for sufficiently small values $W_{12}$ there appears a term proportional to $W_{12}\Delta_1\Delta_2$ in the free energy expression (\ref{e8}) which hints at the same conclusion (in the corresponding domain $\Delta_1$ depends on $W_{12}$ very weakly). Note also that intrinsic phase transition of the stronger band is affected in a qualitatively different way, for instance, $\Delta_1(T_{\mathrm{c}1})\sim\sqrt{1-\frac{T_{\mathrm{c}1}}{T_\mathrm{c}}}\sim |W_{12}|$ and $\frac{\partial\Delta_1(T_{\mathrm{c}1})}{\partial |W_{12}|}\sim |W_{12}|^0$ for vanishing interband interaction. That pairing rather shifts critical point from $T_{\mathrm{c}1}$ to $T_\mathrm{c}$, but not smears it as in the case of weaker band.

An impact of interband interaction on the weaker-superconductivity component established should also manifest itself in the behaviour of non-thermodynamic properties, e.g. spatial coherency. Note that due to interband coupling there appear critical and non-critical channels in the spatial variations \cite{babaev1,litak2012,ord1} (and temporal relaxation \cite{ord2}) of gap fluctuations. Each channel is characterised by its own correlation length, and both of them participate in the coherency properties of each band involved. However, in the vicinity of $T_{\mathrm{c}2}$ the non-critical channel can be neglected for the weaker band \cite{babaev1,GL}. At that, the correlation length of the critical channel should change as $h^{-\frac{1}{3}}$ under "applied external field" $h\sim |W_{12}|$, similarly to the weaker-band healing length \cite{shanenko}. Thus, the peculiarities of spatial coherency in the weaker band also follow at $T_{\mathrm{c}2}$ the Landau theory of phase transitions with interband pairing in the role of external field. 

Unlike other source fields, e.g. external electric or magnetic field in ferroelectric or ferromagnetic systems, correspondingly, interband interaction constant in two-band superconductors is not easily tunable parameter. However, it was illustrated that the coupling between gap order parameters can be varied by changing the size of superconducting structure \cite{bluhm}, doping \cite{MgAlB1}, pressure \cite{pressure} etc.

\section{Smearing of a weaker-band phase transition}
\begin{figure*}
\resizebox{1.1\columnwidth}{!}{
\includegraphics{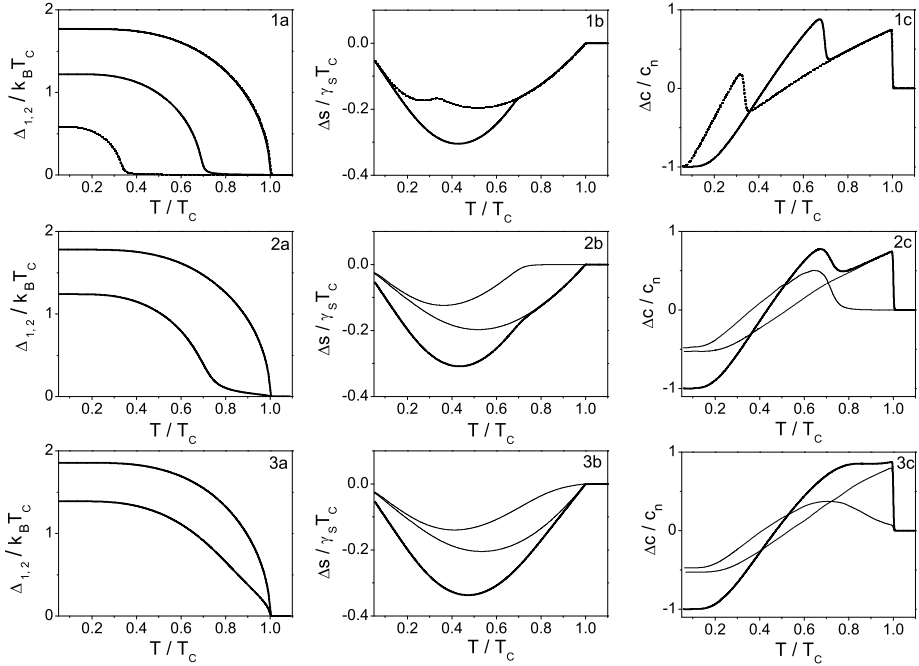}}\vspace{-1cm}
\caption{The plots of the gaps (a), excess specific entropy (b) and normalized excess specific heat capacity (c) vs reduced temperature for
$W_{12}=0.0001\mathrm{\ eV}\cdot\mathrm{cell}$ (1),
$W_{12}=0.001\mathrm{\ eV}\cdot\mathrm{cell}$ (2) and
$W_{12}=0.01\mathrm{\ eV}\cdot\mathrm{cell}$ (3). In the first row
of figures the dashed curves denote the corresponding quantities for
slightly weaker coupling in the second band $W_{22}=0.25\mathrm{\
eV}\cdot\mathrm{cell}$ ($T_{\mathrm{c2}}=0.33T_{\mathrm{c1}}$). In Fig. (1a) the larger gaps flow together for
the constants $W_{22}$ considered. In the second and third rows
of figures the thin curves denote band contributions to the excess
entropy and specific heat capacity.}\label{f2}
\end{figure*}

The presence of  the weaker-band superconducting instability smeared slightly by interband interaction affects drastically the thermodynamics of a two-gap superconductor. Fig. \ref{f2} shows the changes of the gaps, excess entropy and heat capacity with intra- and interband interactions calculated numerically on the basis of Eqs. (\ref{e5})-(\ref{e6}). By turning interband coupling on, the interband proximity effect takes place, i.e. the vanishing of the smaller gap $\Delta_2$ at $T_{\mathrm{c}2}$ changes into inflection near $T_{\mathrm{c}+}$ with simultaneous vanishing of the band superconductivity at $T_\mathrm{c}=T_{\mathrm{c}-}$. That inflection indicates the crossover from active to passive regime of the weaker band and it disappears as $|W_{12}|$ exceeds some value.

For tiny interband coupling the band contributions to entropy and heat capacity (see panels 2b,c in Fig. \ref{f2}) develop considerably in different temperature regions below $T_\mathrm{c}$. Their additive impact results in the substantial increment for net entropy and the non-monotonicity for net heat capacity in the vicinity of $T_{\mathrm{c}+}$. The latter was observed also experimentally \cite{MgB2,FeSe2,cuprates1}. For sufficiently strong interband pairings the memory about weaker-band criticality disappears.

Interestingly, weak interband coupling together with the certain configuration of intraband interaction channels result in the local maximum of entropy in superconducting state, see the first row of panels in Fig. \ref{f2}. In this case $\frac{\Delta c}{c_\mathrm{n}}\sim\frac{\ud\Delta s}{\ud T}$ crosses zero below $T_\mathrm{c}$ three times. Similar effect was revealed in Cu1234  \cite{cuprates2}. We predict also same feature for FeSe$_{1-x}$ \cite{FeSe1} approximately at $3.4$ K.

Usually, external field rounds the singularities related to the phase transition anomaly as well as locates them into positions shifted differently in the vicinity of critical temperature. In a two-gap system interband interaction acts in the same way with respect to weaker-superconductivity component. However, the presence of the band with stronger superconductivity obscures the interrelations between atypical peculiarities seen in Fig. \ref{f2} near $T_{\mathrm{c}2}$. We analyse that problem by calculating inflection points for smaller gap and for temperature derivatives of free energy.

In the case of extremely small interband couplings the solution of Eq. (\ref{e11}) is characterized by the following inflection point
\begin{equation}\label{e19}
T_\mathrm{inf}^{\Delta_2}\approx T_{\mathrm{c}+}-\frac{\Delta_{2+}^{\prime\prime}}{\Delta_{2+}^\mathrm{I\!I\!I}}= T_{\mathrm{c}+}\Bigg(1+\frac{9y_+^{\{\frac{1}{3}\}2}}{2A_+^{\{0\}2}}+\frac{3B_+^{\{1\}}}{2A_+^{\{0\}}y_+^{\{\frac{1}{3}\}}}\Bigg),
\end{equation}
The temperature $T_\mathrm{inf}^{\Delta_2}$ contains the lowest order correction to $T_{\mathrm{c}+}$ proportional to
$|W_{12}|^\frac{4}{3}$. It proves that the smaller gap as a function of temperature inflects in the vicinity of $T_{\mathrm{c}+}$ if interband interaction is extremely weak.

\begin{figure}
\resizebox{1.1\columnwidth}{!}{\hspace{0cm}\includegraphics{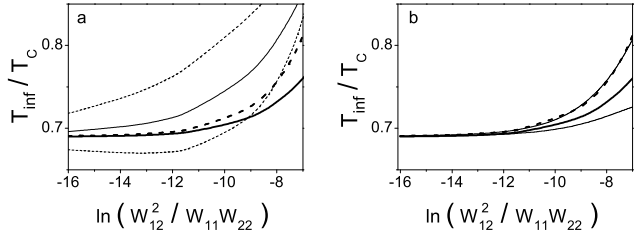}}\vspace{-6cm}
\caption{(a) The evolution of inflection points for $\Delta f^\prime$ (two thin dashed curves for maximum and minimum of $\frac{\Delta c}{c_\mathrm{n}}$), $\Delta f^{\prime\prime}$ (thin solid curve), $\Delta f^\mathrm{I\!I\!I}$ (thick solid curve) and for $\Delta_2$ (thick dashed curve) with interband interaction constant near $T_{\mathrm{c}2}$. (b) Inflection points for $\Delta_2$ (thick dashed curve) and for $\Delta f^\mathrm{I\!I\!I}$ (thick solid curve) compared to the approximations given by the analytic formulas (\ref{e19}) and (\ref{e20}) (thin solid curves). In both figures we calculate inflection points on the basis of free energy (\ref{e5}) and gap equations (\ref{e6}).}\label{f3}
\end{figure}

The dependence $\Delta_2(T)$ defines uniquely the temperature behaviour of free energy (\ref{e8}) which we construct as the Taylor series in the vicinity of $T_{\mathrm{c}+}$. Since sharp non-monotonic behaviour of heat capacity in Fig. \ref{f2} is of particular interest, we search for the zeros of the third temperature derivative $\Delta f^\mathrm{I\!I\!I}(T)$ near $T_{\mathrm{c}+}$. Our calculation indicates that functions $\Delta f^\mathrm{I\!I\!I}(T)$ and $\Delta f^\mathrm{I\!V}(T)$ vanish remarkably farther off $T_{\mathrm{c}+}$ as opposed to $\Delta f^\mathrm{V}(T)$. In other words, neither extrema nor inflection points of $\frac{\Delta c}{c_\mathrm{n}}$ appear as close to $T_{\mathrm{c}+}$ as $T_\mathrm{inf}^{\Delta_2}$ for tiny interband interactions. This conclusion is confirmed also by numeric analysis of free energy (\ref{e5}) and gap equations (\ref{e6}), see Fig. \ref{f3}a. 

The vanishing of $\Delta f^\mathrm{V}(T)$ sufficiently close to $T_{\mathrm{c}+}$ for tiny interband pairings points to an inflection point of $\Delta f^\mathrm{I\!I\!I}(T)$, namely,
\begin{equation}\label{e20}
T_\mathrm{inf}^{\Delta
f^\mathrm{I\!I\!I}}\approx T_{\mathrm{c}+}-\frac{\Delta
f_+^\mathrm{V}}{\Delta
f_+^\mathrm{V\!I}}=
T_{\mathrm{c}+}\Bigg(1+\frac{3B_+^{\{1\}}}{2A_+^{\{0\}}y_+^{\{\frac{1}{3}\}}}\Bigg).
\end{equation}
That inflection stems from the behaviour of the weaker-band contribution to $\Delta f^\mathrm{I\!I\!I}(T)$. Expressions (\ref{e19}) and (\ref{e20}) approximate values found numerically strikingly well (see Fig. \ref{f3}b). Moreover, these formulas evidences the matching of inflection points for smaller gap and for third derivative of free energy, when $T_{\mathrm{c}1,2}$ are close (the second term in Eq. (\ref{e19}) can be neglected). Since large difference between $T_{\mathrm{c}1,2}$ may be compensated by the weakness of interband interaction, this matching may be more general than analytic approach we use. Numerics presented support the conclusion.

\section{Conclusions}The growing family of milti-component superconductors requires deep insight into the processes introduced by inter-component couplings. Various examples show that corresponding physics becomes non-trivial and it is hardly deducible from the superposition of non-interacting superconducting subsystems. He we reported new understanding of the impact of interband pair-transfer interaction on the formation of two-band superconducting order. Namely, the latter pairing was shown to play the role of external field associated with smaller-gap order parameter. The finding can be of importance for the interpretation of the experimental data in the superconducting materials with tiny interband coupling, e.g. FeSe$_{1-x}$, V$_3$Si etc.

\ack
We thank Egor Babaev for valuable discussions. The study was supported by the European Union through the European Regional Development Fund (Centre of Excellence "Mesosystems: Theory and Applications", TK114) and by the Estonian Science Foundation (Grant No 8991).

\section*{References}
\bibliography{2bandTherm}

\end{document}